\documentclass[a4paper,twocolumn,
english,aps,prb,floatfix,showpacs]{revtex4}
\usepackage[T1]{fontenc}
\usepackage[latin1]{inputenc}
\usepackage{amsmath}
\usepackage{babel}
\usepackage{graphics}
\usepackage{amssymb}


\begin{document}
\title
{Quantum Diffusion and Localization in Disordered Electronic Systems}
\author {P. R. \surname{Wells Jr.}}

\email{wells@if.ufrj.br}

\author {J. \surname{d$^\prime$Albuquerque~e~Castro}}

\email{jcastro@if.ufrj.br}

\author {S. L. A. \surname{de~Queiroz}}

\email{sldq@if.ufrj.br}

\affiliation{Instituto de F\'\i sica, Universidade Federal do
Rio de Janeiro, Caixa Postal 68528, 21941-972
Rio de Janeiro RJ, Brazil}

\date{\today}

\begin{abstract}
The diffusion of electronic wave packets in one-dimensional
systems with on-site, binary disorder is numerically investigated 
within the framework of a single-band tight-binding model.
Fractal properties
are incorporated by assuming that the distribution of distances
$\ell$ between consecutive impurities obeys a power law, 
$P(\ell) \sim \ell^{-\alpha}$. For suitable ranges of $\alpha$,
one  finds system-wide anomalous diffusion. 
Asymmetric diffusion effects are introduced through the
application of an external electric field,
leading to results similar to those observed in the case of
photogenerated electron-hole plasmas in tilted InP/InGaAs/InP quantum
wells.
\end{abstract}

 
\maketitle
 
\section{Introduction} 
\label{intro}
%
%
The work reported in this paper is motivated by the experimental observation 
of asymmetric diffusion of electron-hole plasmas in 
semiconductor quantum wells~\cite{monte00,monte02,monte06}, as well as by 
attempts at a theoretical explanation of such results, on the basis of models for
fractal diffusion~\cite{monte00,ch98}. In Ref.~\onlinecite{monte00}, 
electron-hole plasmas were photogenerated in intrinsic InP/InGaAs/InP single quantum 
wells, and their diffusion was observed through measurements of
photoluminescence intensity profiles. Two types of heterostructures were examined:
the first was grown in a direction normal to the $[001]$ crystallographic axis, 
and the second was grown in a direction tilted by $2^\circ$ relative to that
of the first, toward the $[111]$ axis. 
The experimental results were as follows: in the former structure the plasma
undergoes Gaussian (symmetric) diffusion, while in the latter the diffusion is 
asymmetric.
The authors of Ref.~\onlinecite{monte00} showed that diffusion data for the
tilted heterostructure can be fitted by an asymmetric L\'evy distribution, previously
obtained~\cite{ch98} as a solution to a (one-dimensional) generalized Fokker-Planck 
equation with distinct right-- and left-- diffusion coefficients, and fractional
derivatives. Furthermore, it was argued that the fractional character of
the diffusion is connected to fractal properties of the medium,
namely a power-law roughness distribution of the
interfaces that delimit the quantum wells. 
Further photoluminescence studies
suggest that the carrier diffusion properties are indeed very sensitive
both to interface roughness, and to the presence of finite-width 
terraces~\cite{monte02}. By varying both optical excitation intensities and
temperature, a more detailed picture was found for the anomalous diffusion phenomena 
taking place, including the likely existence of Auger recombination, which in its
turn might be induced by fractal interface morphology~\cite{monte06}.      

The relationship between fractional diffusion and the absence of a
characteristic length scale (the latter being a basic property of
fractals) is well-established, and has been extensively
investigated~\cite{mk00,z02,calvo}.
 
Here we study a model system along the lines of Anderson's 
picture~\cite{anderson:58,prl42,kbb85,mckk93} for electron diffusion in disordered media. 
Alternatively to the theoretical approach adopted in Refs.~\onlinecite{monte00,ch98},
whose starting point was a generalized Fokker-Planck equation for (classical) particles, 
we write a tight-binding Hamiltonian for a (quantum) particle, which evolves 
according to the rules of quantum mechanics. The fractal features are incorporated
by a specific spatial distribution of on-site disorder, to be described below.

Bearing in mind that the electron-hole pairs undergoing anomalous 
diffusion in the experimental systems are essentially quantum-mechanical
objects, our treatment is expected to reflect the basic wave-like 
properties of such entities. The link between the quantum properties 
underlying localization, and those classical features brought to fore
in classical approaches to diffusion is not well understood, although it 
has been discussed in the recent past (see, e.g., Ref.~\onlinecite{gg04}).

Our main concern is to  provide an independent check of whether the 
connection between fractal properties of a disordered medium, and the behavior
of  particles diffusing in it, which has been proposed in the classical 
context, is 
robust enough to survive the translation to a quantum-mechanical picture.
Therefore, 
we shall not aim at making specific comparisons of our results to the experiments
described in Refs.~\onlinecite{monte00,monte02,monte06}.

In Subsection~\ref{subsec:2a}, we introduce the tight-binding hamiltonian 
and the respective disorder distributions to be investigated, along with
a brief description of our calculational procedures. 
In Subsection~\ref{subsec:2.b} results are given for the time evolution 
of the width of wavefunction packets against time, as well as for the
corresponding instantaneous profiles. Subsection~\ref{sec:II.3} deals
with the introduction of an external bias, in order to investigate
the effects of anisotropy. Finally, in Sec.~\ref{conc}, concluding remarks
are made.

\section
{Model system and Results}
\label{sec:2}

\subsection{Model}
\label{subsec:2a}

In order to simplify the calculational framework, we restrict our investigation to
one-dimensional systems.
We consider the one-electron, tight-binding hamiltonian for a 
single-orbital linear chain with nearest neighbor hopping,
\begin{equation}
H =\sum_{n}\,| n \rangle\,\epsilon_n \,\langle n |\,-\,\gamma
\,\sum_{n,m}\, | n \rangle\,\langle m | \, ,
\label{tbh}
\end{equation}
where $| n \rangle$ is an orbital with self-energy $\epsilon_n$, $m=n
\pm 1$, and  $\gamma$ is the hopping energy, assumed to be constant and
positive. 

We study binary models of disordered alloys, in which two 
different orbitals $A$ (host atoms) and $B$ (substitutional impurities) coexist,
with respective self-energies $\epsilon_A$ and  
$\epsilon_B$, and overall concentrations $1-p_0$ and $p_0$.
The relevant energy in the present context 
is $\epsilon \equiv \epsilon_B - \epsilon_A$, which measures
the impurity-host mismatch. 

If the impurity locations are uncorrelated, (a case to be denoted as the
{\it random binary model}, or random for short) the distances $\ell$ 
between any 
two consecutive impurities have
an exponential
distribution, i.e.,
\begin{equation}
P(\ell) =\frac{p_0}{1-p_0}\,(1-p_0)^\ell \sim \exp (- \lambda \ell)\quad 
(\ell=1,2,\dots)\,,
\label{pell}
\end{equation}
where $\lambda =\lambda(p_0)=-\ln (1-p_0)$ is the inverse decay length 
(in lattice parameter units) that sets a scale for the typical $B-B$
distance, i.e., a cutoff 
dictating the largest allowable distances. Note that the {\em average} 
impurity-impurity distance is $p_0^{-1}$, which approaches $\lambda^{-1}$
only in the low-concentration limit $p_0 \ll 1$ where continuum and
discrete-lattice descriptions become equivalent. For the random model,
host-impurity  ($A-B$) duality means that our investigation can be 
restricted to $0 \leq p_0 \leq 1/2$.

We introduce fractal features by assuming the distribution of $\ell$ to decay 
with a power law:
\begin{equation} 
P(\ell) \sim \ell^{-\alpha} \ , 
\label{alpha} 
\end{equation} 
where $\alpha$ is a characteristic exponent. 

Distributions similar to Eq.~(\ref{alpha}) have been extensively studied
in the literature of fractal-based point processes~\cite{lt92,lt93,ltbk};
in that context, they refer to the probabilities of occurrence of interevent 
(time) separations.
In such cases, most of the interest focuses on the characteristics of the
associated power spectrum, which turns out to exhibit $1/f^\alpha$ noise
properties. Here, by contrast, the fractal features of the real-space  
impurity distribution are of interest only inasmuch as they are the background
against which we simulate the quantum-mechanical evolution of electronic
wavepackets.

The distribution given by Eq.~(\ref{alpha}) is normalizable only for $\alpha > 1$; 
its mean is finite only for $\alpha > 2$ , and its variance is finite 
only for $\alpha > 3$. Nevertheless, as discussed at length in 
Subsec.~\ref{subsec:2.b}, it is a physically sensible choice to adopt a 
system-wide normalization which enables one to
consider $1 < \alpha \leq 2$, for (large but finite) fractal systems.
In this {\it fractal binary 
model}, or fractal, for short, $\ell$ lacks a typical scale. Here, all 
distances are allowed, on account of the slowly-decaying distribution 
tail. Thus we expect the resulting system to exhibit fractal properties, 
provided that it is large enough. The correlation between impurity 
positions, implied by Eq.~(\ref{alpha}), destroys strict $A-B$ duality.

Comparison between overall concentrations in the two cases proceeds by 
matching the respective average distances between impurities, which are 
$p_0^{-1}$
for the former model [~as remarked in connection with 
Eq.~(\ref{pell})~]
and $\zeta (\alpha-1)/\zeta(\alpha)$ 
for the latter, where $\zeta (\alpha) =\sum_{n=1}^{\infty} n^{-\alpha}$ is 
Riemann's zeta function. For example, $p_0=0.4 \leftrightarrow \alpha 
\approx 2.34$.
Clearly, in the thermodynamic limit any 
$\alpha \leq 2$ corresponds to vanishing $B$ concentrations in the 
fractal. However, as pointed out above, for large but finite systems 
one can still get nontrivial results for 
$\alpha \leq 2$, once suitable normalization considerations are taken into 
account 
[~see Fig.~\ref{p0vsalpha} below, for a full illustration of the $p_0$--$\,\alpha$
correspondence, both in the thermodynamic limit and for finite systems; see also 
Eq.~(\ref{zeta_L})~].

We study the dynamics of wave packets~\cite{pla194,zhong} in random and
fractal linear chains.
The time evolution of the amplitudes $\psi_n  (t) = \langle n | \psi (t) \rangle$ 
is determined by the Schr\"odinger equation:
\begin{equation} 
i \, \dot{\psi}_n \, = \, \varepsilon_n \, \psi_n \, - \, \psi_{n - 1} \, - \, \psi_{n + 
1} \, \, ,
\label{diffeq}
\end{equation} 
where $\varepsilon_n = \epsilon_n/\gamma$ is a dimensionless parameter, and time is 
given in units of $\hbar/\gamma$.

We start with Gaussian wave packets:
\begin{equation} 
\psi_n (0) = C \, \exp \left[ \frac{-(n-n_0)^2}{2 
\sigma_0^2} \right]\ ,
\label{initpac}
\end{equation} 
where  $n_0$ denotes the initial position of the centroid,
and $\sigma_0$ the initial width, and $C$ is a normalization constant.

Configurational averages are taken in all the calculations presented in this work.

\subsection{Variances and wave fronts}
\label{subsec:2.b}

We analyze and compare the properties of the two kinds of systems just defined.

We have numerically determined the time evolution of the (ensemble-) 
averaged wave-packets in one-dimensional systems of sizes $L = 
10000$, 
with free ends. The number of 
realizations incorporated in the ensemble averages is $M = 1000$
or, in some cases, larger.

The calculational method is based on numerical integration of 
Eq.~(\ref{diffeq}) with initial condition given by Eq.~(\ref{initpac}), 
via a fifth-order Runge-Kutta code. We compared results for fixed 
realizations of disorder, obtained with time steps of $0.1$ and $0.01$, 
and checked that they are indistinguishable for all practical purposes. 
Thus, we set the former value in all calculations described here.

The integration is taken up to times not longer than enough for the packet 
to reach the chain's ends, in order to avoid reflection effects. The 
actual hopping-rate of an electron in a pure system modeled by 
Eq.~(\ref{diffeq}) is one atom per time unit. Considering that diffusion 
on a disordered chain is hampered by impurities, in order to be safe we 
set the upper time limit as $L/2$, for a packet initially spread over a 
few sites around the center of an $L$-atom chain. We have used 
$\sigma_0=5.0$, that is, the initial packets are rather localized.

As usual, we study diffusional behavior through investigation of the time 
evolution of the ensemble-averaged second moment (variance) of the 
particle probability distribution, $\langle (\Delta x)^2\,\rangle$. 
Power-law dependencies, $\langle (\Delta x)^2\,\rangle\ \sim t^{2\beta}$, 
with $\beta=0$, $1/2$, and $1$ characterize localized, diffusive, and 
``ballistic'' regimes respectively (the latter to be interpreted here
in the sense that, even though there is no center-of-mass motion in the
absence of an external bias, the
packet width varies linearly in time if each of its
Fourier components travels unhindered). 

Initially we consider the random model, so as to provide a benchmark against
which to compare results for the fractal case. 

\begin{figure}
{\centering \resizebox*{3.3in}{!}{\includegraphics*{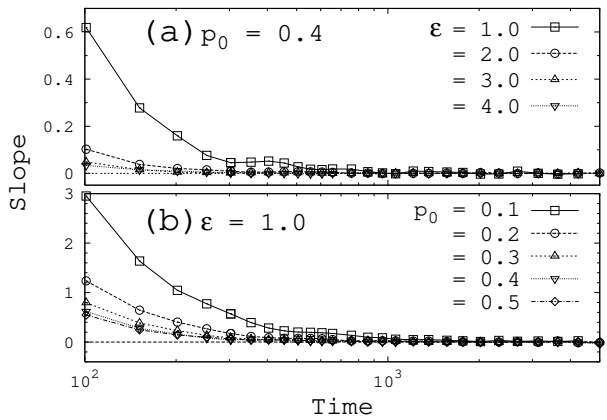}}}
\caption{
Plots of slopes (time derivatives) $d\,\langle (\Delta x)^2\,\rangle/dt$ 
of variances of particle probability distribution 
against time in the random system.  In (a) $\varepsilon=1.0$ for several 
$p_0$, while in (b), $p_0=0.4$ for several $\varepsilon$.  $L=10\,000$ and
$M=10^3$ samples each. 
}
\label{random-var} 
\end{figure}

By scanning the values of $\varepsilon \equiv 
\epsilon/\gamma$ for the dimensionless impurity-host self-energy 
mismatch, as well as those of $p_0$, along suitable intervals, one finds 
the same overall picture, namely apparent ballistic behavior at first, 
followed by a continuous, smooth crossover towards localization,
i.e. the variances $\langle (\Delta x)^2\,\rangle$ eventually approach 
saturation. 
To illustrate this, in Figure \ref{random-var} we show 
the evolution of the slopes $d\,\langle (\Delta x)^2\,\rangle/dt$ against 
time, in the random system. One can see that saturation behavior (zero
slope) always obtains, albeit at rates which depend on $\varepsilon$
and $p_0$. 

We have found no simple scaling picture, from which a data collapse plot 
could be derived. Even for fixed $\varepsilon$, where for each $p_0$ one 
has the characteristic inter-impurity distance $\lambda^{-1}(p_0)$, this 
quantity does not translate directly on to a scaling length. In this case, the 
saturation value of $\langle (\Delta x)^2\,\rangle$ varies approximately 
as $p_0^{-1.35}$ (for $p_0 \leq 0.5$).

\begin{figure}
{\centering \resizebox*{3.3in}{!}{\includegraphics*{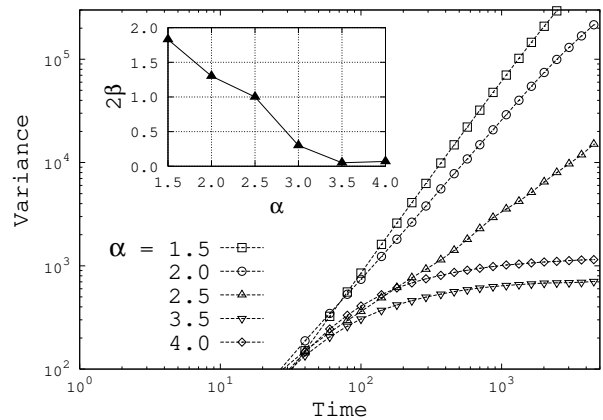}}}
\caption{
Double-logarithmic plots of variance of particle probability distribution 
against time in the fractal system, for $\varepsilon=1.0$ and several 
$\alpha$ (see text for explanation of $\alpha \leq 2$ data). $L=10\,000$ and
$M=5 \times 10^3$ samples each [$\,$except $\alpha=1.5$ for which
$M=10^3\,$]. Inset: asymptotic characteristic exponent $2\beta$
as a function of $\alpha$, for data on the main diagram (see text).
} 
\label{fractal-var}
\end{figure}

\begin{figure}
{\centering \resizebox*{3.3in}{!}{\includegraphics*{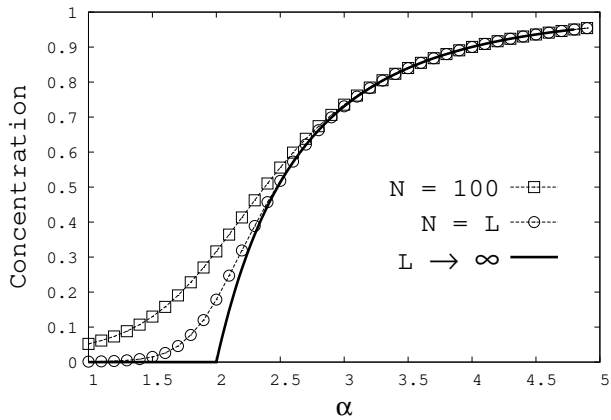}}}
\caption{
Effective average concentration $p_0$ of impurity sites in the fractal 
model, against $\alpha$. $L \to \infty$ refers to  
Eq.~({\protect{\ref{zeta_L}}}) in such limit, while points are for 
$L=10\,000$, with normalization by $N$ [$N=L$ is system-wide normalization
as in Eq.~({\protect{\ref{zeta_L}}})]; see text.
}
\label{p0vsalpha}
\end{figure}

We now turn to the fractal model, for which representative  
results are shown in Fig. \ref{fractal-var}. These are for samples whose distributions 
$P(\ell)$ are normalized with respect to the actual system size $L$, i.e.,
\begin{equation}
P(\ell) =\frac{\ell^{-\alpha}}{\zeta_L(\alpha)}\ ,\quad \zeta_L(\alpha) \equiv \sum_{n=1}^L \frac{1}{n^\alpha}\ .
\label{zeta_L}
\end{equation}
Such a system-wide normalization is indeed consistent with the inclusion-exclusion 
principle. This can be seen by recalling that, starting from a $B$ atom at $x=0$ 
and following along the chain, the statement that, e.g., the first
$B$ atom occurs at $x=\ell$ encompasses all (mutually exclusive)
$2^{L-\ell}$ possible arrangements of  $A$'s  and $B$'s, for $x > \ell$,
and with {\em only} $A$'s for $1 \leq x
\leq \ell-1$. 
With the above normalization, even for $\alpha \leq 2$ one has, on
average, a non-vanishing 
(though in such case rather small) impurity concentration. This is illustrated in 
Fig.~\ref{p0vsalpha}.

The larger the system, the 
individual probabilities for some given distance $\ell$ become 
comparatively smaller, while the occurrence of larger distances becomes 
possible, although with low probability. We find that for larger 
systems, the fractal environment makes it easier for the packet to diffuse 
than the random one, for the same (effective) concentration of impurities.
Of course, here one is only considering diffusion along distances 
of the order of the system's length, $L$. 
The crucial difference, relative to the random model, is that since there is no
intrinsic length scale in the fractal system, the single length introduced
by the normalization in Eq.~(\ref{zeta_L}) coincides with the system size.
Thus, although the diffusive behavior observed here must always be
regarded as an apparent regime, it may extend to rather long distances.

The occurrence of large impurity-impurity distances in this system may 
produce individual realizations of disorder, comprising regions that are
effectively pure. The 
result is that the eigenstates in the system may be extended, or 
semi-extended. Averaging over many realizations would produce an ensemble 
of wave packets composed of localized and extended states, even though on 
average the system has a finite concentration of impurities. Least-squares 
fits of $10^2 < t < 5 \times 10^3$ data of
Figure~\ref{fractal-var} to a 
single power-law form produce an exponent $2\beta=1.82(1)$, $1.30(1)$, and 
$1.04(1)$ respectively for $\alpha=1.5$, $2.0$, and $2.5$. Thus one finds 
an effective anomalous diffusion regime, which has no counterpart in the 
random disorder model.

On the other hand, the $\alpha=3.5$ results, for example, show that even
in the fractal 
model, one can have disorder so strong that localization 
(within the system's limits)
is statistically the only possible outcome. For this case, the
corresponding effective 
impurity concentration, as defined above, is $\approx 0.84$, which
(for the random model) would be equivalent to $p_0=0.16$, by using
the $p_0 \leftrightarrow 1-p_0$ duality valid for the latter type of
disorder.
One can infer from Fig.~\ref{random-var}~(b) that the
behavior of the $p_0=0.16$ random system is indeed similar to that of the
$\alpha=3.5$ fractal case, namely localization setting in for $t
\approx 10^3$
[~we have also checked that the actual saturation value of 
$\langle (\Delta x)^2\,\rangle$, which is $\approx 6 \times10^2$
for $\alpha=3.5$ (fractal), falls between the respective ones
for $p_0=0.1$ and $0.2$ (random)~].

As regards 
$\alpha=4.0$ data, for which one has the equivalent $p_0=0.90$
from  Fig.~\ref{p0vsalpha}, comparison is to be
made to the $p_0=0.10$ curve of Fig.~\ref{random-var}~(b). Again, the
agreement is very good, as far as general trends are concerned: 
increasing $\alpha$ in this range turns out to produce a longer
localization length.

The reason why the fractal model behaves similarly to the random one
for large $\alpha$, while
it certainly does not do so for $\alpha \leq 2.5$, is as follows.
From Eqs.~(\ref{alpha}) and~(\ref{zeta_L}), one can work out that the
probability distribution for distances between consecutive $A$ atoms is
\begin{equation}
P_A(\ell) \sim \exp \left[a(\alpha)\,\ell\right] ,\quad a(\alpha)= \ln
P(1)\ ,
\label{eq:pa}
\end{equation}
where $P(1)=\zeta_L^{-1}(\alpha)$ is the probability for unit distance
between $B$ (impurity) atoms. For $\alpha \leq 2.5$
(corresponding to $p_0 \leq 0.5$), the $B$ atoms are in the minority, 
so they indeed play the role of impurities in an otherwise
pure $A$ system. As seen above, the fractal properties associated to the 
power-law $B-B$ distance distribution, Eq.~(\ref{alpha}), give rise to the 
consequent diffusion-like behavior.
At larger $\alpha$ ($p_0 > 0.5$), the $A$ atoms
are now in the minority. Physically, the traveling electrons
are only sensitive to the existence of two  distinct values of on-site 
energies, thus the effective "impurity" label will be assigned in practice to 
the species which occurs less frequently (in this case, $A$ atoms,
whose distance distribution, Eq.~(\ref{eq:pa}), is qualitatively
the same as in the random model, thus bringing about localization).
The characteristic length, $\left[\ln \zeta_L(\alpha)\right]^{-1}$,
increases with increasing $\alpha$, thus explaining the trend
mentioned above. 

In practice, one would expect the region $0.4 \lesssim p_0 \lesssim 0.6$,   
i.e., $2.35 \lesssim \alpha \lesssim 2.65$, to behave as a crossover 
region.
This is because, for the finite systems under study, one needs a clear 
majority of one species over the other to be statistically established 
while still within distances shorter than system size.

\begin{figure}
{\centering \resizebox*{3.3in}{!}{\includegraphics*{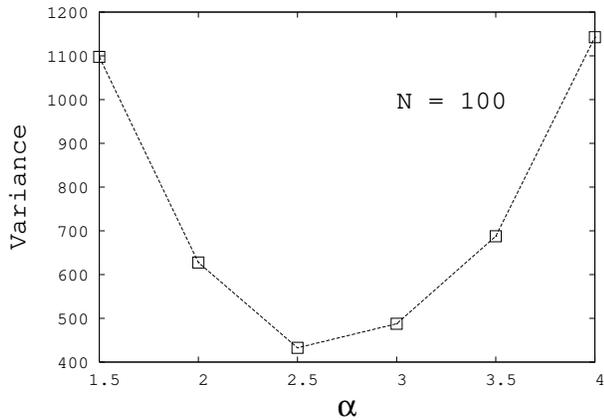}}}
\caption{
Asymptotic values of variance of particle probability distribution  in the
fractal system, with $P(\ell)=0$ for $\ell>100$, and the corresponding normalization (see 
text), for $\varepsilon=1.0$ and  several $\alpha$. 
$L=10\,000$ and $M=10^3$ samples each. 
}
\label{f-cutoff}
\end{figure}

Further evidence that effective diffusion-like behavior is linked to the 
absence of a typical scale in fractal systems can be derived as follows. 
If, instead of normalizing $P(\ell)$ by system size $L$ as in 
Eq.~(\ref{zeta_L}), we take a fixed $N < L$ as the upper limit (see in Fig.~\ref{p0vsalpha}
how this affects the effective impurity concentration),
a length scale equal to $N$ is introduced, even though the variation of $P(\ell)$ 
against $\ell$ is still described by a power law. Results for $N=100$, and 
assorted values of $\alpha$ are shown in Fig.~\ref{f-cutoff}.
Note that while the localization length decreases with increasing $\alpha
\leq 2.5$, the trend is reversed for $\alpha > 2.5$. This is similar in
nature to the $p_0 \leftrightarrow 1-p_0$ duality observed in the random
model.

Finally, we examine the actual shapes of particle probability densities. 
For the localized packets in the random case,
extended exponentials, $|\psi (x)|^2 \sim \exp(-b|x|^\phi)$, provide
reasonably 
good fits to the region within $20$ to $30$ sites from the peak, with 
$1.5 \lesssim \phi \lesssim 1.8$.
At larger distances, the probability density decay is somewhat slower than that.
Fits for $30 \lesssim x \lesssim 200$ give $\phi$ in the $0.2 - 0.6$
range. 

\begin{figure}
{\centering \resizebox*{3.3in}{!}{\includegraphics*{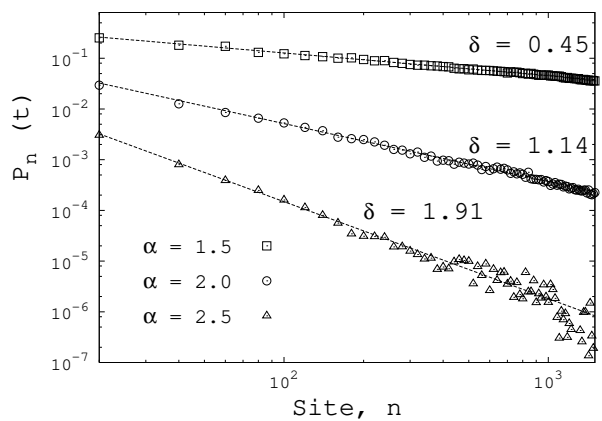}}}
\caption{
Double-logarithmic plots of particle probability distribution functions 
against position, at $t=5000$, for assorted values of $\alpha$ in the 
(apparent) diffusive regime. $\varepsilon=1.0$, $L=10\,000$, and
$M=10^3$ samples. Lines indicate least-squares fits to a power-law form,
$P(x) \sim x^{-\delta}$. 
See text for a discussion of suitable normalization
considerations. 
}
\label{prof-fits}
\end{figure}

On the other hand, for the fractal cases with apparent diffusion, one can get
power-law fits extending to two or more decades of distance, as shown in
Fig.~\ref{prof-fits}. 
Of course, the power-law behavior exhibited in the Figure is expected to
hold only within the system's finite limits, as is the case for all
features of the apparent diffusion regime (recall that our whole study
is conducted for times not longer than enough for the wavepacket to reach the
chain's ends, in order to avoid reflection effects). Thus, for instance,
although data for $\alpha=1.5$ (corresponding to power-law decay with 
$\delta=0.45$) would be strictly non-normalizable if such behavior 
extended to arbitrarily long distances, one must keep in mind that 
(within the present context) the wavepacket amplitude will
fall to zero before reaching the chain's ends.  
Localized packets for fractal disorder, i.e., 
$\alpha \gtrsim 3.0$ (not shown), seem to behave in an intermediate way
between power-law and extended exponential decay.  

\subsection{Application of a bias}
\label{sec:II.3}

We search for an effect that, in the present quantum-mechanical formulation,
would be the equivalent to distinct values for left- and right- diffusion
coefficients in the Fokker-Planck approach of Refs.~\onlinecite{monte00,ch98}.
The simplest source of such anisotropy  is an electric field.

The effect of applying a uniform electric field to the system is incorporated by 
the inclusion of the term
\begin{equation}
H_{bias} = \sum_n | \, n \rangle e \, E \, n \, \langle n |
\label{field}
\end{equation}
in the Hamiltonian, where $e$ is the electronic charge, and $E$ is the field intensity. 
The corresponding Schr\"odinger equation is
\begin{equation}
i \, \dot{\psi}_n \, = \, ( \varepsilon_n + f_0 \, n) \, \psi_n \, - \, \psi_{n - 1} \, 
- \, \psi_{n + 1} \, \, ,
\label{diffeq-bias}
\end{equation}
where $f_0 = e \, E / \gamma$ is a dimensionless bias intensity. 

On a pure system, application of a bias produces a drift of the entire wave packet, as 
its centroid moves through the lattice, with diffusion occurring with respect to the 
center-of-mass reference frame. However, due to the unequal effect of the bias on the various 
Fourier components of the packet, this diffusion may be asymmetric. Also, one must
be aware of Bloch oscillations, which confine the wave packet in a region of space, thus 
producing oscillating behavior from application of a static electric field~\cite{fl92}.

In order to prevent the effect of Bloch oscillations from distorting the
diffusive behavior which is our main concern, here we use $f_0 = 1.0 \times 10^{-3}$.
Then, elementary considerations show that for a packet starting at the center of
a chain with $L=10\,000$, speed reversal will only set in at $t \lesssim 4 \times 10^3$,
giving one a rather broad window of observation. 

In a disordered system, the application of a bias gives rise to dynamical 
localization. This is related to Bloch oscillations. The difference is 
that a part of the packet diffuses away and performs oscillations, while the other 
part remains localized close to the origin. Such asymmetric diffusion behavior may 
thus be viewed as the coexistence of two distinct regimes.
\begin{figure}
{\centering \resizebox*{3.3in}{!}{\includegraphics*{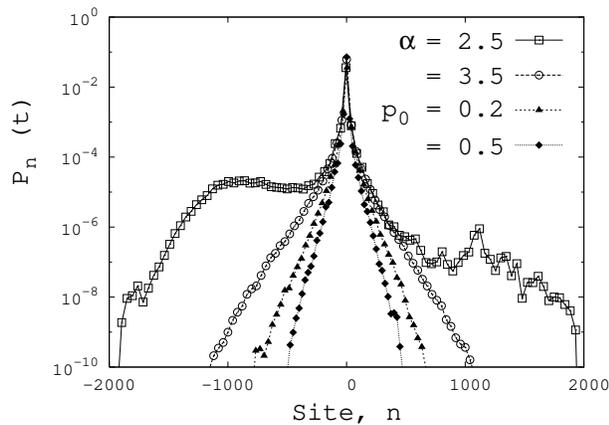}}}
\caption{
Particle probability distribution against position, for $t=1000$, $f_0=1.0 \times 
10^{-3}$ 
for $\varepsilon=1.0$.  $L=10\,000$ and $M=10^3$ samples each [$\,$except
$\alpha=2.5$ for which  $M=5\times 10^3\,$]. Open symbols: fractal systems; 
full symbols: random systems.
}
\label{prof-bias}
\end{figure}

The results in Fig.~\ref{prof-bias} show 
that the diffusing portion of the averaged wave packet behaves as if the system were 
pure;
its centroid coincides with that of a corresponding wave packet in a pure system. 
One can see that it is only for the fractal case with apparent
diffusion ($\alpha=2.5$) that a significant 
portion of the particle probability distribution is
pure-system-like. As remarked above, for such values of $\alpha$ one has
coexistence between individual disorder realizations with 
effectively delocalized eigenstates, and others in which localization
occurs. The relative height of the corresponding peaks in
Fig.~\ref{prof-bias} 
shows that the latter are much more frequent than the former. 
\newline 

\section{Discussion and Conclusions} 
\label{conc}

We have introduced a model system for the incorporation of on-site fractal disorder in 
the one-electron diffusion problem. As remarked above, the distribution of impurity-impurity 
distances given in Eq.~(\ref{alpha}) implies that disorder is in fact correlated. It is 
well known that features such as long-range hopping~\cite{esc88,prb69} and/or correlated 
disorder~\cite{dwp90,pw91,sve94,bel99,im05}  may have 
strong effects (including the occurrence of a metal-insulator transition even in 
one-dimensional systems), an observation which is confirmed here by the comparison of 
fractally-disordered and uncorrelated random electronic systems.

As regards comparison with experimental results, the wavefront profiles
depicted in Fig.~\ref{prof-bias} show a connection between
anomalous diffusion of tight-binding electrons and
fractal properties of the underlying medium. Furthermore, we have seen 
that, in the present model, asymmetric profiles arise from coexistence
of ballistic and localized states upon application of an external bias.
Such coexistence results from the fact that a subset of
ensemble realizations of the (scale-free) disorder are, in fact, almost
pure.
Whether the same explanation holds for the experimentally-observed
profiles is not certain at the moment, though one might conceivably
propose ways to test it on available samples.

\begin{acknowledgments}   

We thank A. S. Chaves and F. A. B. F. de Moura
for interesting conversations. 
This research  was  partially supported by
the Brazilian agencies CAPES, CNPq,
FAPERJ (Grant No. E26--100.604/2007), and
Instituto do Mil\^enio de Nanotecnologia--MCT.
\end{acknowledgments}

\end{document}